# Localized degenerate solutions to the massless Dirac and Weyl equations

Georgios N. Tsigaridas[1,*], Aristides I. Kechriniotis[2], Christos A. Tsonos[2] and Konstantinos K. Delibasis[3]

[1]Department of Physics, School of Applied Mathematical and Physical Sciences, National Technical University of Athens, GR-15772 Zografou Athens, Greece

[2]Department of Physics, University of Thessaly, GR-35100 Lamia, Greece

[3]Department of Computer Science and Biomedical Informatics, University of Thessaly, GR-35131 Lamia, Greece

[*]Corresponding Author. E-mail: gtsig@mail.ntua.gr

**Abstract**

In this article we present a general class of localized degenerate solutions to the massless Dirac and Weyl equations, which can also describe particles, or systems of particles, with varying energy and spin along their direction of motion. Another interesting characteristic of these solutions is that they remain unaltered in a wide range of electromagnetic 4-potentials and fields, which are analytically calculated. In addition, we propose a new method for spatially separating Weyl particles based on their helicity and direction of motion using appropriate magnetic fields, given in explicit form.

**Keywords**: Dirac equation, Weyl equations, Dirac particles, Weyl particles, Massless particles, Localized solutions, Degenerate solutions, Electromagnetic 4-potentials, Electromagnetic fields

1. Introduction

In a recent article [1], we have shown that all solutions to the Weyl equations are degenerate, in the sense that they remain unaltered under a wide range of electromagnetic 4-potentials and fields, which are explicitly calculated. In the same article, we have also shown that under certain conditions, e.g. massless particles, the solutions to the Dirac equation can also become degenerate, corresponding to an infinite number of electromagnetic 4-potentials and fields. In [2] we have found degenerate solutions to the Dirac equation in the case of massive particles, associated with quantum tunneling. Furthermore, in [3] we have provided some general forms of degenerate solutions corresponding to massless particles. In [4] we have shown that Weyl particles can become localized even in the absence of electromagnetic fields and that the localization of the particles can be easily controlled through simple electric fields. Based on these results, we have proposed a novel device for controlling the flow of information at exceptionally high rates using Weyl fermions [5]. Additionally, in [6] we have provided degenerate wave-like solutions to the Dirac equation for massive particles and in [7] we describe a general method for obtaining degenerate solutions to the Dirac and Weyl equations, giving also some hints regarding the experimental detection of degenerate states. Finally, in [8] we provide a comprehensive review on this interesting topic.

These results are expected to be particularly useful for the deeper understanding of the behavior of materials supporting massless Dirac and/or Weyl particles, such as graphene sheets [9-16] and Weyl semimetals [17-35], opening new pathways regarding their technological applications.

In the present work we provide a general class of localized degenerate solutions to the massless Dirac and Weyl equations, being also able to describe particles, or systems of particles, with varying energy and spin along their direction of motion. These solutions could be associated with virtual particles appearing in materials supporting massless charged particles [9-35]. They may also be related to spontaneous wave-function localization [36] and to a new theoretical model suggesting that all elementary particles could be represented using fractionally charged Weyl spinors [37]. Another interesting characteristic of these solutions is that they are degenerate, corresponding to a wide variety of electromagnetic 4-potentials and fields, which are analytically calculated. Additionally, in section 3, we propose a new method for spatially separating Weyl particles according to their helicity and direction of motion, using appropriate magnetic fields.

## 2. Localized solutions to the massless Dirac equation

Let us consider the Dirac equation in the form of [1, 38, 39]

$$i\gamma^\mu \partial_\mu \Psi + a_\mu \gamma^\mu \Psi - m\Psi = 0, \qquad (2.1)$$

where $\gamma^\mu$ are the four contravariant gamma matrices in the Dirac representation

$$\gamma^0 = \begin{pmatrix} \sigma^0 & 0 \\ 0 & -\sigma^0 \end{pmatrix}, \quad \gamma^\mu = \begin{pmatrix} 0 & \sigma^\mu \\ -\sigma^\mu & 0 \end{pmatrix}, \quad \mu = 1,2,3 \qquad (2.2)$$

and $\sigma^\mu$ are the standard Pauli matrices given by given by the following formulae

$$\sigma^0 = \begin{pmatrix} 1 & 0 \\ 0 & 1 \end{pmatrix} \quad \sigma^1 = \begin{pmatrix} 0 & 1 \\ 1 & 0 \end{pmatrix} \quad \sigma^2 = \begin{pmatrix} 0 & -i \\ i & 0 \end{pmatrix} \quad \sigma^3 = \begin{pmatrix} 1 & 0 \\ 0 & -1 \end{pmatrix}. \qquad (2.3)$$

Additionally, $m$ is the mass of the particles and $a_\mu = qA_\mu$, where $q$ is the charge of the particles and $A_\mu$ is the electromagnetic 4-potential. It should also be noted that Dirac equation is expressed in natural units, where $\hbar = c = 1$.

Starting from the well-known free-particle solutions [1, 38, 39], we have found that all spinors of the form

$$\Psi_p = \left[ f(w) \begin{pmatrix} \cos\left(\frac{\theta}{2}\right) \\ e^{i\varphi} \sin\left(\frac{\theta}{2}\right) \\ \cos\left(\frac{\theta}{2}\right) \\ e^{i\varphi} \sin\left(\frac{\theta}{2}\right) \end{pmatrix} + g(w) \begin{pmatrix} -\sin\left(\frac{\theta}{2}\right) \\ e^{i\varphi} \cos\left(\frac{\theta}{2}\right) \\ \sin\left(\frac{\theta}{2}\right) \\ -e^{i\varphi} \cos\left(\frac{\theta}{2}\right) \end{pmatrix} \right] \exp(ih(w)) \qquad (2.4)$$

and

$$\Psi_a = \left[ f(w) \begin{pmatrix} \sin\left(\frac{\theta}{2}\right) \\ -e^{i\varphi} \cos\left(\frac{\theta}{2}\right) \\ -\sin\left(\frac{\theta}{2}\right) \\ e^{i\varphi} \cos\left(\frac{\theta}{2}\right) \end{pmatrix} + g(w) \begin{pmatrix} \cos\left(\frac{\theta}{2}\right) \\ e^{i\varphi} \sin\left(\frac{\theta}{2}\right) \\ \cos\left(\frac{\theta}{2}\right) \\ e^{i\varphi} \sin\left(\frac{\theta}{2}\right) \end{pmatrix} \right] \exp(ih(w)) \qquad (2.5)$$

where $f(w)$, $g(w)$, $h(w)$ are real functions of $w = \sin\theta\cos\varphi\, x + \sin\theta\sin\varphi\, y + \cos\theta\, z - t$, are solutions of the massless $(m=0)$ Dirac equation for zero 4-potential, describing particles and antiparticles respectively. Obviously, the functions $f(w)$, $g(w)$ must be square integrable so that the spinors (2.4) and (2.5) are normalizable. In the case that the functions $f(w)$, $g(w)$ are constants and $h(w) = Ew$, where $E$ is a real positive constant corresponding to the energy of the particles, the above spinors describe free massless Dirac particles and antiparticles respectively, moving with energy $E$ along a direction in space defined by the polar angle $\theta$ and the azimuthal angle $\varphi$ in spherical coordinates [1, 38, 39].

It should be noted that the spinors (2.4) and (2.5) contain three arbitrary functions and consequently they provide a huge flexibility for describing almost any desired state. For example, supposing that the arbitrary functions $f(w)$, $g(w)$ are of the form $B + A\exp\left(-k(w-w_0)^2\right)$ where $B$, $A$, $k$, $w_0$ are real constants with $k$ being positive, the above spinors describe localized solutions around the position $w = w_0$, or $\sin\theta\cos\varphi\, x + \sin\theta\sin\varphi\, y + \cos\theta\, z = t + w_0$. Obviously, in the case that $B = 0$ these solutions become fully localized around the position $w = w_0$. Furthermore, since the functions $f(w)$, $g(w)$ are arbitrary, they can also be linear combinations of Gaussian functions in the form $A_1 \exp(-k_1(w-w_{01})^2) + A_2 \exp(-k_2(w-w_{02})^2) + ...$, representing particles localized at many different positions along their direction of motion. Additionally, the localization positions are not stationary but move at the speed of light along the direction of motion of the particles. Obviously, in the case of particles existing in materials as collective quasiparticle excitations, the speed of light is not $c = 3 \times 10^8 \, m/s$, but it is equal to the maximum speed supported by the materials, which can be considerably lower than $c$.

As far as the function $h(w)$ is concerned, the derivative $dh/dw$ corresponds to the energy of the particles. Thus, spinors (2.4) and (2.5) can also describe particles, or system of particles with varying energy along their direction of motion. For example, if we suppose that $h(w) = \sqrt{\pi/\lambda}(E_0/2) \text{erf}(\sqrt{\lambda}(w-w_0))$, where $E_0$, $\lambda$ are real positive constants and erf is the error function, then the energy of the particles also follows a Gaussian distribution of the form $E(w) = E_0 \exp(-\lambda(w-w_0)^2)$. Thus, particles have non-zero energy only around the position $w = w_0$, while their energy tends to zero as $|w-w_0| \to \infty$. Obviously, particles described by spinors (2.4) and (2.5) can also be both localized and with variable energy. For example, supposing that $f(w) = g(w) = A \exp(-k(w-w_0)^2)$ and $h(w) = \sqrt{\pi/\lambda}(E_0/2)\text{erf}(\sqrt{\lambda}(w-w_0))$, particles described by the above spinors exist only around the position $w = w_0$ and vanish as $|w-w_0| \to \infty$. Consequently, they could be interpreted as virtual particles, appearing in materials supporting massless charged particles, such as graphene sheets [9-16] and Weyl semimetals [17-35]. They could also be related to spontaneous wave-function localization in the framework of relativistic quantum mechanics [36].

In figure 1 we depict the real part of the first component of a spinor of the form (2.4) or (2.5) around the position $w = w_0$ in the case that $f(w) = g(w) = A \exp(-k(w-w_0)^2)$ and $h(w) = \sqrt{\pi/\lambda}(E_0/2)\text{erf}(\sqrt{\lambda}(w-w_0))$.
Obviously, it has the form of a wave with variable amplitude, corresponding to the localization of the particles and variable frequency, corresponding to the varying energy of the particles.

Another interesting remark is that the spin components of the particles corresponding to the spinors (2.4) and (2.5), as calculated through the following formulae [6, 7, 38, 40]

$$S_{x,p} = -S_{x,a} = \frac{i}{2} \Psi^\dagger \gamma^2 \gamma^3 \Psi = \frac{1}{2} \sin\theta \cos\varphi \frac{f(w)^2 - g(w)^2}{f(w)^2 + g(w)^2} \tag{2.6}$$

$$S_{y,p} = -S_{y,a} = \frac{i}{2}\Psi^\dagger \gamma^3 \gamma^1 \Psi = \frac{1}{2}\sin\theta \sin\varphi \frac{f(w)^2 - g(w)^2}{f(w)^2 + g(w)^2} \tag{2.7}$$

$$S_{z,p} = -S_{z,a} = \frac{i}{2}\Psi^\dagger \gamma^1 \gamma^2 \Psi = \frac{1}{2}\cos\theta \frac{f(w)^2 - g(w)^2}{f(w)^2 + g(w)^2} \tag{2.8}$$

where $\Psi^\dagger$ is the Hermitian adjoint of $\Psi$, are generally functions of $w$. Here, the subscripts $p$, $a$ correspond to particles and antiparticles respectively, which obviously have spin opposite to each other. The total spin of the particles and antiparticles is defined through the formula

$$S_p = S_a = \sqrt{S_x^2 + S_y^2 + S_z^2} = \frac{1}{2}\left|\frac{f(w)^2 - g(w)^2}{f(w)^2 + g(w)^2}\right| \tag{2.9}$$

which is also a function of $w$.

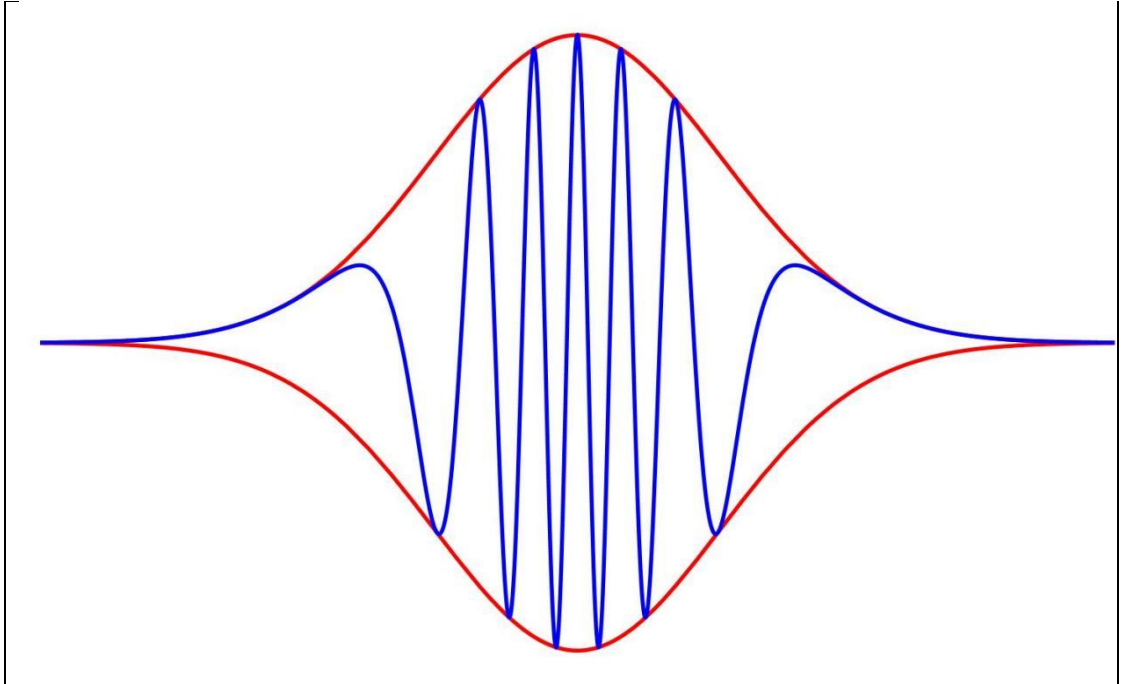

**Figure 1**: The real part of the first component of a spinor of the form (2.4) or (2.5) around the position $w = w_0$ in the case that $f(w) = g(w) = A\exp\left(-k(w-w_0)^2\right)$ and $h(w) = \sqrt{\pi/\lambda}\,(E_0/2)\,\text{erf}\left(\sqrt{\lambda}(w-w_0)\right)$.

Thus, the spinors (2.4) and (2.5) may generally describe systems of particles, or antiparticles, where the relative orientation of their spins varies along their direction of motion. However, if $f(w) = 0$ or $g(w) = 0$, then, according to Eq. (2.9), the spin of the particles becomes constant and equal to $1/2$. Consequently, in this special case, the spinors (2.4) and (2.5) may describe single massless fermions or antifermions.

Furthermore, if $f(w) = g(w)$, the spin of the particles becomes equal to zero. Thus, in this case, the spinors (2.4) and (2.5) may describe pairs of particles or antiparticles with spin opposite to each other.

Moreover, it should be emphasized that the spinors (2.4) and (2.5) are degenerate. Consequently, according to Theorem 5.4 in [1], these spinors are solutions to the massless Dirac equation not only for the initial zero 4-potential $a_\mu$, but also for a wide range of additional 4-potentials $b_\mu$, given by the formula [1]

$$b_\mu = a_\mu + s(\mathbf{r},t)\theta_\mu \qquad (2.10)$$

where $s(\mathbf{r},t)$ is an arbitrary real function of the spatial coordinates and time and

$$(\theta_0, \theta_1, \theta_2, \theta_3) = \left(1, -\frac{\Psi^T \gamma^0 \gamma^1 \gamma^2 \Psi}{\Psi^T \gamma^2 \Psi}, -\frac{\Psi^T \gamma^0 \Psi}{\Psi^T \gamma^2 \Psi}, \frac{\Psi^T \gamma^0 \gamma^2 \gamma^3 \Psi}{\Psi^T \gamma^2 \Psi}\right)$$
$$= (1, -\sin\theta\cos\varphi, -\sin\theta\sin\varphi, -\cos\theta) \qquad (2.11)$$

Here, $\Psi^T$ is the transpose of $\Psi$. It should also be noted that the 4-potentials $b_\mu$ are the same both for particles and antiparticles. The electromagnetic fields corresponding to the 4-potentials $b_\mu$ can be easily calculated [41] through the following formula

$$\mathbf{E} = -\nabla U - \frac{\partial \mathbf{A}}{\partial t}, \quad \mathbf{B} = \nabla \times \mathbf{A} \qquad (2.12)$$

where $U = a_0/q$ is the electric potential and $\mathbf{A} = -(1/q)(a_1\hat{\mathbf{x}} + a_2\hat{\mathbf{y}} + a_3\hat{\mathbf{z}})$ is the magnetic vector potential. Here, $\hat{\mathbf{x}}$, $\hat{\mathbf{y}}$, $\hat{\mathbf{z}}$ are the unit vectors along the x, y, z axes respectively. The choice of the minus sign in the definition of the magnetic potential is related to choice of the Dirac matrices $\gamma^\mu$ used in this article.

Through Eq. (2.12) we derive the electromagnetic fields corresponding to the 4-potentials $b_\mu$:

$$\mathbf{E} = -\frac{1}{q}(s_t \sin\theta\cos\varphi + s_x)\hat{\mathbf{x}} - \frac{1}{q}(s_t \sin\theta\sin\varphi + s_y)\hat{\mathbf{y}} - \frac{1}{q}(s_t \cos\theta + s_z)\hat{\mathbf{z}} \qquad (2.13)$$

$$\mathbf{B} = \frac{1}{q}(s_y \cos\theta - s_z \sin\theta\sin\varphi)\hat{\mathbf{x}} - \frac{1}{q}(s_x \cos\theta - s_z \sin\theta\cos\varphi)\hat{\mathbf{y}}$$
$$+ \frac{1}{q}\sin\theta(s_x \sin\varphi - s_y \cos\varphi)\hat{\mathbf{z}} \qquad (2.14)$$

Here, the subscripts $x$, $y$, $z$, $t$ denote partial differentiation with respect to the corresponding coordinate, e.g. $s_x = \partial s/\partial x$, $s_y = \partial s/\partial y$, etc. Thus, the spinors (2.4) and (2.5) are solutions to the massless Dirac equation not only in zero electromagnetic field, but also in the wide range of electromagnetic fields given by equations (2.13) and (2.14).

### 3. Localized solutions to the Weyl equations and a proposed method for spatially controlling and separating Weyl particles

The above results can be easily extended to the case of Weyl equations. Specifically, starting from the well-known free-particle solutions [1, 38, 39], we have found that the spinors

$$\psi_p = f(w) \begin{pmatrix} \cos\left(\frac{\theta}{2}\right) \\ e^{i\varphi} \sin\left(\frac{\theta}{2}\right) \end{pmatrix} \exp(ih(w)) \qquad (3.1)$$

and

$$\psi_n = f(w) \begin{pmatrix} -\sin\left(\frac{\theta}{2}\right) \\ e^{i\varphi} \cos\left(\frac{\theta}{2}\right) \end{pmatrix} \exp(ih(w)) \qquad (3.2)$$

are solutions to the Weyl equations

$$i\sigma^\mu \partial_\mu \psi + a_\mu \sigma^\mu \psi = 0 \qquad (3.3)$$

and

$$i\sigma^\mu \partial_\mu \psi - 2i\sigma^0 \partial_0 \psi + a_\mu \sigma^\mu \psi - 2a_0 \sigma^0 \psi = 0 \qquad (3.4)$$

corresponding to particles with positive and negative helicity respectively, for zero 4-potential. Here, $f(w)$, $h(w)$ are real functions of $w = \sin\theta\cos\varphi\, x + \sin\theta\sin\varphi\, y + \cos\theta\, z - t$ and $\theta$, $\varphi$ are real constants corresponding to the polar and azimuthal angle respectively, defining the propagation direction of the particles in spherical coordinates. Furthermore, the function $f(w)$ should be square integrable so that the spinors (3.1) and (3.2) are normalizable.

As in the case of the massless Dirac equation, the spinors (3.1) and (3.2) can also describe localized particles and particles with varying energy along their direction of motion. For example, setting $f(w) = A\exp\left(-k(w-w_0)^2\right)$ and $h(w) = \sqrt{\pi/\lambda}\,(E_0/2)\mathrm{erf}\left(\sqrt{\lambda}(w-w_0)\right)$, the spinors (3.1) and (3.2) describe particles with variable energy given by the formula $E = E_0 \exp\left(-\lambda(w-w_0)^2\right)$, which are also localized around the position $w = w_0$. In this case, the spinors (3.1) and (3.2) can be visualized as localized waves with variable frequency, shown in figure 1. However, contrary to the case of massless Dirac particles, Weyl particles have constant spin equal to $1/2$, being either parallel or antiparallel to their direction of motion in the cases of positive and negative helicity, respectively.

Here, it should also be mentioned that the localized Weyl solutions described above may also be related to a new theoretical model suggesting that fractionally charged Weyl spinors could form the basis of elementary particles, including all known fermions and bosons, even predicting a new type of massive neutral particles that could be a dark matter candidate [37].

Furthermore, according to Theorem 3.1 in [1], all Weyl solutions are degenerate, corresponding to an infinite number of 4-potentials given by the formula

$$b_\mu = a_\mu + s(\mathbf{r},t)\varphi_\mu \tag{3.5}$$

where

$$(\varphi_0, \varphi_1, \varphi_2, \varphi_3) = \left(1, \mp\frac{\psi^\dagger \sigma^1 \psi}{\psi^\dagger \psi}, \mp\frac{\psi^\dagger \sigma^2 \psi}{\psi^\dagger \psi}, \mp\frac{\psi^\dagger \sigma^3 \psi}{\psi^\dagger \psi}\right) \tag{3.6}$$

and $s(\mathbf{r},t)$ is an arbitrary real function of the spatial coordinates and time. Here, $\psi^\dagger$ is the Hermitian adjoint spinor of $\psi$. It should also be noted that the minus and plus sign in Eq. (3.6) corresponds to the cases of positive and negative helicity respectively. Thus, the spinors (3.1) and (3.2) are also solutions to the Weyl equations for the 4-potentials

$$(b_0, b_1, b_2, b_3) = (1, -\sin\theta\cos\varphi, -\sin\theta\sin\varphi, -\cos\theta)s(\mathbf{r},t) \tag{3.7}$$

which are the same both for particles with positive and negative helicity. The above 4-potentials $b_\mu$ correspond to the electromagnetic fields given be equations (2.13) and (2.14), as in the case of massless Dirac particles.

Additionally, we have found that the spinors

$$\psi_p = p(x,y) f(z-t) \begin{pmatrix} 1 \\ 0 \end{pmatrix} \exp(ih(z-t)) \tag{3.8}$$

and

$$\psi_n = p(x,y) f(z-t) \begin{pmatrix} 0 \\ 1 \end{pmatrix} \exp(ih(z-t)) \tag{3.9}$$

where $p(x,y)$ is a real function of the transverse coordinates $x$, $y$ and $f(z-t)$, $h(z-t)$ are real functions of $z-t$, are also solutions to the Weyl equations (3.3) and (3.4) for the 4-potentials

$$(a_0, a_1, a_2, a_3)_p = (0, p_y/p, -p_x/p, 0) \tag{3.10}$$

and

$$(a_0, a_1, a_2, a_3)_n = (0, -p_y/p, p_x/p, 0) \tag{3.11}$$

respectively. Here, it should be noted that these solutions are a generalization of the spinors presented in section 3 in [5], corresponding to particles moving along the $+z$ direction with energy $E$ and a specific transverse spatial distribution, defined by the function $p(x, y)$, e.g. a Gaussian or super-Gaussian function. It should also be noted that the functions $p(x, y)$ and $f(z-t)$ must be square integrable so that the spinors (3.8) and (3.9) are normalizable.

Obviously, these solutions can also describe localized particles and particles with varying energy along their direction of motion. For example, if we suppose that $f(z-t) = A \exp\left(-k(z-t-z_0)^2\right)$ and $h(z-t) = \sqrt{\pi/\lambda}\,(E_0/2)\,\mathrm{erf}\left(\sqrt{\lambda}(z-t-z_0)\right)$, where $z_0$ is a real constant, the spinors (3.8) and (3.9) describe particles localized around the position $z = z_0 + t$ with variable energy given by the formula $E = E_0 \exp\left(-\lambda(z-t-z_0)^2\right)$.

Furthermore, according to equations (3.5) and (3.6), the full set of 4-potentials $b_\mu$ which are associated with these solutions is given by the formulae

$$(b_0, b_1, b_2, b_3)_p = (0, p_y/p, -p_x/p, 0) + (1, 0, 0, -1) s(\mathbf{r}, t) \tag{3.12}$$

and

$$(b_0, b_1, b_2, b_3)_n = (0, -p_y/p, p_x/p, 0) + (1, 0, 0, -1) s(\mathbf{r}, t) \tag{3.13}$$

corresponding to the following electromagnetic fields:

$$\mathbf{E}_p = -\frac{1}{q}\left(s_x \hat{\mathbf{x}} + s_y \hat{\mathbf{y}} + (s_z + s_t)\hat{\mathbf{z}}\right) \tag{3.14}$$

$$\mathbf{B}_p = \frac{1}{q}\left(s_y \hat{\mathbf{x}} - s_x \hat{\mathbf{y}}\right) - \frac{1}{q}\frac{1}{p^2}\left(p_x^2 + p_y^2 - p(p_{xx} + p_{yy})\right)\hat{\mathbf{z}} \tag{3.15}$$

in the case of particles with positive helicity and

$$\mathbf{E}_n = -\frac{1}{q}\left(s_x \hat{\mathbf{x}} - s_y \hat{\mathbf{y}} - (s_z + s_t)\hat{\mathbf{z}}\right) = \mathbf{E}_p \tag{3.16}$$

$$\mathbf{B}_n = \frac{1}{q}\left(s_y \hat{\mathbf{x}} - s_x \hat{\mathbf{y}}\right) + \frac{1}{q}\frac{1}{p^2}\left(p_x^2 + p_y^2 - p(p_{xx} + p_{yy})\right)\hat{\mathbf{z}} \tag{3.17}$$

in the case of particles with negative helicity.

Similarly, the spinors

$$\psi_p = p(x,y) f(z+t) \begin{pmatrix} 0 \\ 1 \end{pmatrix} \exp(ih(z+t)) \qquad (3.18)$$

and

$$\psi_n = p(x,y) f(z+t) \begin{pmatrix} 1 \\ 0 \end{pmatrix} \exp(ih(z+t)) \qquad (3.19)$$

where $f(z+t)$, $h(z+t)$ are real functions of $z+t$, are solutions to the Weyl equations (3.3) and (3.4) for the 4-potentials

$$(a_0, a_1, a_2, a_3)_p = (0, -p_y/p, p_x/p, 0) \qquad (3.20)$$

and

$$(a_0, a_1, a_2, a_3)_n = (0, p_y/p, -p_x/p, 0) \qquad (3.21)$$

respectively. These spinors correspond to particles moving along the $-z$ direction with a specific transverse spatial distribution, defined by the function $p(x,y)$. As in the previous cases, they can also describe localized particles and particles with varying energy along their direction of motion. For example, supposing that $f(z+t) = A\exp(-k(z+t-z_0)^2)$ and $h(z+t) = \sqrt{\pi/\lambda}(E_0/2)\mathrm{erf}(\sqrt{\lambda}(z+t-z_0))$ the spinors (3.18) and (3.19) describe particles localized around the position $z = z_0 - t$ with variable energy given by the formula $E = E_0 \exp(-\lambda(z+t-z_0)^2)$. Obviously, the functions $p(x,y)$ and $f(z+t)$ must be square integrable so that the spinors (3.18) and (3.19) are normalizable.

The full set of 4-potentials connected to these solutions is

$$(b_0, b_1, b_2, b_3)_p = (0, -p_y/p, p_x/p, 0) + (1,0,0,1) s(\mathbf{r},t) \qquad (3.22)$$

and

$$(b_0, b_1, b_2, b_3)_n = (0, p_y/p, -p_x/p, 0) + (1,0,0,1) s(\mathbf{r},t) \qquad (3.23)$$

corresponding to the following electromagnetic fields:

$$\mathbf{E}_p = -\frac{1}{q}\left(s_x \hat{\mathbf{x}} + s_y \hat{\mathbf{y}} + (s_z - s_t)\hat{\mathbf{z}}\right) \qquad (3.24)$$

$$\mathbf{B}_p = \frac{1}{q}\left(-s_y \hat{\mathbf{x}} + s_x \hat{\mathbf{y}}\right) + \frac{1}{q}\frac{1}{p^2}\left(p_x^2 + p_y^2 - p(p_{xx} + p_{yy})\right)\hat{\mathbf{z}} \qquad (3.25)$$

and

$$\mathbf{E}_n = -\frac{1}{q}\left(s_x\hat{\mathbf{x}} + s_y\hat{\mathbf{y}} + (s_z - s_t)\hat{\mathbf{z}}\right) = \mathbf{E}_p \tag{3.26}$$

$$\mathbf{B}_n = \frac{1}{q}\left(-s_y\hat{\mathbf{x}} + s_x\hat{\mathbf{y}}\right) - \frac{1}{q}\frac{1}{p^2}\left(p_x^2 + p_y^2 - p(p_{xx} + p_{yy})\right)\hat{\mathbf{z}} \tag{3.27}$$

in the case of particles with positive and negative helicity, respectively.

Another interesting remark is that, if the free function $s(\mathbf{r},t)$ is equal to zero, then only the $z-$component of the magnetic field survives and the electromagnetic fields corresponding to these spinors obtain the simple form

$$\mathbf{B}_{\mp} = \mp\frac{1}{q}\frac{1}{p^2}\left(p_x^2 + p_y^2 - p(p_{xx} + p_{yy})\right)\hat{\mathbf{z}}, \tag{3.28}$$

where the minus (plus) sign corresponds to particles with positive (negative) helicity moving along the $+z(-z)$ direction or particles with negative (positive) helicity moving along the $-z(+z)$ direction.

Furthermore, if we choose the function $r(x,y) = r_1/p(x,y)$ to describe the transverse spatial distribution of the particles, then the magnetic fields in Eq. (3.28) change sign. Therefore, since $r_1$ is a real constant, the magnetic fields given by Eq. (3.28) could also correspond to particles with spatial distribution given by the function $r(x,y) = r_1/p(x,y)$. In this case, the plus (minus) sign corresponds to particles with positive (negative) helicity moving along the $+z$ direction or particles with negative (positive) helicity moving along the $-z$ direction.

In this way, particles with different helicities and particles moving along opposite directions can become spatially separated. For example, we assume that the distribution $p(x,y)$ is super-Gaussian, i.e.

$$p(x,y) = A\exp\left(-k_1(x-x_0)^{2n_1} - k_2(y-y_0)^{2n_2}\right) \tag{3.29}$$

where $A$, $x_0$, $y_0$, $k_1$, $k_2$ are real constants with $k_1$, $k_2$ being positive and $n_1$, $n_2$ are positive integers. In this case, the magnetic fields given by Eq. (3.28) take the form

$$\mathbf{B}_{\mp} = \pm\frac{2}{q}\left(k_1 n_1(1-2n_1)(x-x_0)^{2(n_1-1)} + k_2 n_2(1-2n_2)(y-y_0)^{2(n_2-1)}\right)\hat{\mathbf{z}} \tag{3.30}$$

Thus, if we apply the magnetic field $\mathbf{B}_-$ along the $z-$axis, then particles with positive helicity moving along the $+z$ direction and particles with negative helicity moving along the $-z$ direction concentrate around the position $(x,y) = (x_0, y_0)$, as shown in figure 2. On the other hand, particles with negative helicity moving along the $+z$ direction and particles with positive helicity moving along the $-z$ direction follow the inverse distribution and move away from the position $(x,y)=(x_0,y_0)$. Similarly, applying the magnetic field $\mathbf{B}_+$ along the $z-$axis, particles with negative helicity moving along the $+z$ direction and particles with positive helicity moving along the $-z$ direction concentrate around the position $(x,y)=(x_0,y_0)$. On the other hand, particles with positive helicity moving along the $+z$ direction and particles with negative helicity moving along the $-z$ direction, follow the inverse distribution and move away from the position $(x,y)=(x_0,y_0)$. Thus, through an appropriate choice of the applied magnetic field, one could spatially separate Weyl particles with different helicities and moving along opposite directions. Obviously, these results are expected to have interesting applications regarding the control of Weyl fermions in materials supporting these particles, such as Weyl semimetals [17-35].

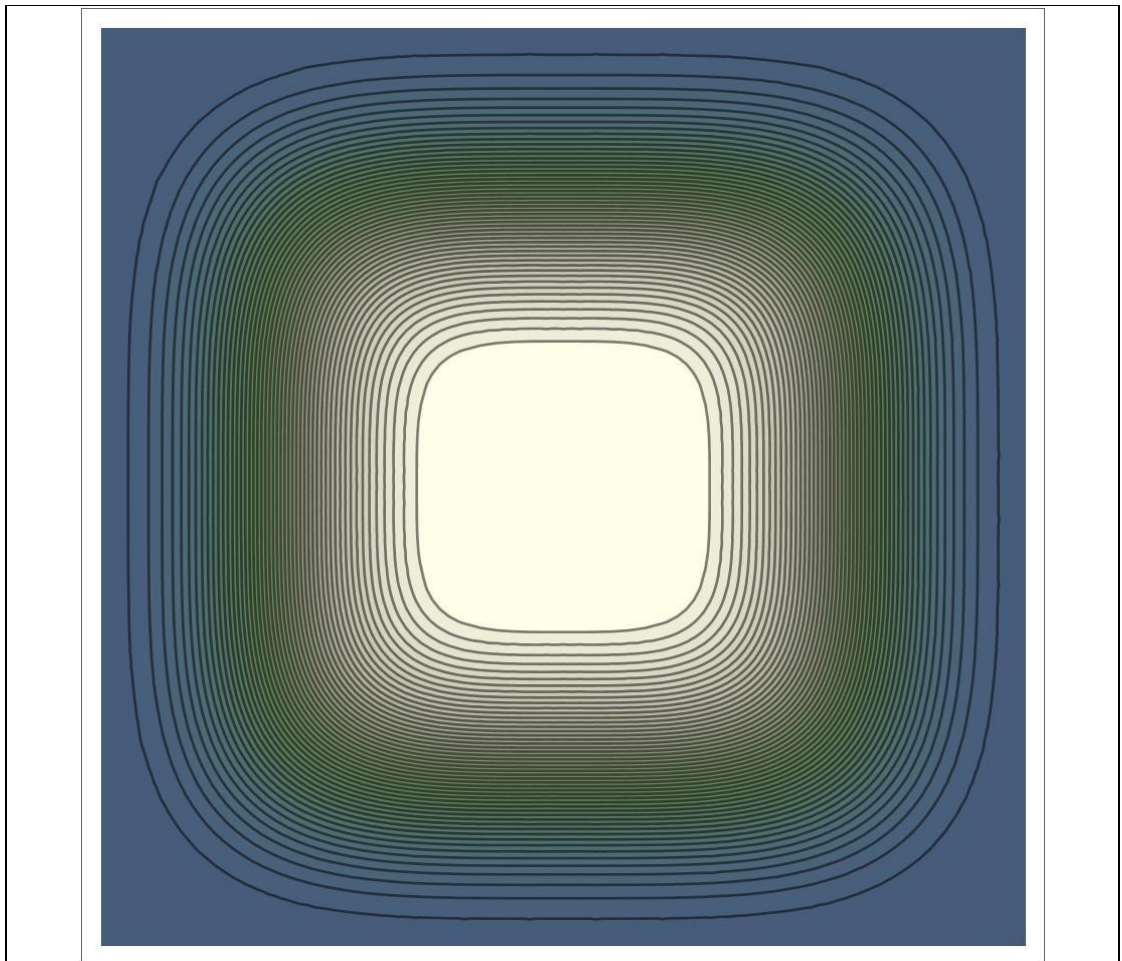

**Figure 2**: Visual representation of the concentration of Weyl particles around the position $(x,y)=(x_0,y_0)$, using appropriate magnetic fields. More details can be found in the main text.

Concluding this section, we should note that localized solutions to the Dirac and Weyl equations have also been presented in [4] and [7]. Specifically, in [4] we have shown that Weyl particles can exist at different states in zero electromagnetic field, either moving as free particles, or being in localized bound states, or even existing in an intermediate state, being bound on the x-y plane, and free along the z-axis. Furthermore, in [7] we have provided a class of degenerate solutions to the Dirac equation, describing localized massive particles in non-zero electromagnetic field.

Finally, we would like to mention that localized solutions to the Dirac equation in zero electromagnetic field have also been presented by other researchers [42-48]. However, most of them consider modified versions of the Dirac equation, e.g. nonlinear Dirac equations in non-zero electromagnetic 4-potentials and fields, with a few exceptions, e.g. [47] and [48], where localized solutions to the standard Dirac equation are provided in zero electromagnetic field. These solutions mainly describe massive particles localized around a specific position in space. On the other hand, our results are more flexible, being able to describe massless particles localized at several positions along their direction of motion and particles with variable energy and spin. In addition, our solutions are degenerate, having the property to remain unaltered in a wide range of electromagnetic 4-potentials and fields, which are calculated analytically. Finally, to the best of our knowledge, this is the first time that localized solutions to the Weyl equations are provided.

## 4. Conclusions

We have presented a general class of localized degenerate solutions to the massless Dirac and Weyl equations which can also describe particles with variable energy and spin, even in zero electromagnetic 4-potential and field. We have also calculated the wide range of electromagnetic 4-potentials and fields corresponding to these solutions. Finally, we have proposed a new method for separating Weyl particles according to their helicity and direction of motion through suitable magnetic fields.

## References

[1] A. I. Kechriniotis, C. A. Tsonos, K. K. Delibasis and G. N. Tsigaridas, On the connection between the solutions to the Dirac and Weyl equations and the corresponding electromagnetic 4-potentials, arXiv:1208.2546 [math-ph], Commun. Theor. Phys. 72 (2020) 045201, DOI: 10.1088/1572-9494/ab690e

[2] G. N. Tsigaridas, A. I. Kechriniotis, C. A. Tsonos and K. K. Delibasis, Degenerate solutions to the Dirac equation for massive particles and their applications in quantum tunneling, arXiv:2010.09844 [quant-ph], Phys. Scr. 96 (2021) 065507, DOI: 10.1088/1402-4896/abf304

[3] G. N. Tsigaridas, A. I. Kechriniotis, C. A. Tsonos and K. K. Delibasis, Degenerate solutions to the massless Dirac and Weyl equations and a proposed method for controlling the quantum state of Weyl particles, arXiv:2010.09846 [quant-ph], Chin. J. Phys. 77, 2324-2332(2022) DOI: 10.1016/j.cjph.2022.04.010

[4] G. N. Tsigaridas, A. I. Kechriniotis, C. A. Tsonos and K. K. Delibasis, On the localization properties of Weyl particles, arXiv:2205.11251 [quant-ph], Ann. Phys. (Berlin) 2200437 (2022) DOI: 10.1002/andp.202200437

[5] G. N. Tsigaridas, A. I. Kechriniotis, C. A. Tsonos and K. K. Delibasis, A proposed device for controlling the flow of information based on Weyl fermions, arXiv:2307.06489 [quant-ph], Sensors 24 (2024) 3361, DOI: 10.3390/s24113361

[6] G. N. Tsigaridas, A. I. Kechriniotis, C. A. Tsonos and K. K. Delibasis, Degenerate wave-like solutions to the Dirac equation for massive particles, arXiv:2209.10933 [quant-ph], EPL 142 (2023) 50001, DOI: 10.1209/0295-5075/acd473

[7] G. N. Tsigaridas, A. I. Kechriniotis, C. A. Tsonos and K. K. Delibasis, A general method for obtaining degenerate solutions to the Dirac and Weyl equations and a discussion on the experimental detection of degenerate states, arXiv:2210.02003 [quant-ph], Ann. Phys. (Berlin) 2200647 (2023) DOI: 10.1002/andp.202200647

[8] G. N. Tsigaridas, A. I. Kechriniotis, C. A. Tsonos and K. K. Delibasis, Degenerate solutions to the Dirac and Weyl equations and their applications, arXiv:2504.00840 [quant-ph]

[9] K. S. Novoselov, A. K. Geim, S. V. Morozov, D. Jiang, M. I. Katsnelson, I. V. Grigorieva, S. V. Dubonos and A. A. Firsov, Two-dimensional gas of massless Dirac fermions in graphene, Nature 438, 197-200 (2005) DOI: 10.1038/nature04233

[10] C. H. Park, L. Yang, Y. W. Son, M. L. Cohen, and S. G. Louie, New Generation of Massless Dirac Fermions in Graphene under External Periodic Potentials, Phys. Rev. Lett. 101, 126804 (2008) DOI: 10.1103/PhysRevLett.101.126804

[11] A. A. Sokolik, A. D. Zabolotskiy and Yu. E. Lozovik, Generalized virial theorem for massless electrons in graphene and other Dirac materials, Phys. Rev. B 93, 195406 (2016) DOI: 10.1103/PhysRevB.93.195406

[12] C. H. Park, L. Yang, Y. W. Son, M. L. Cohen, and S. G. Louie, Anisotropic behaviours of massless Dirac fermions in graphene under periodic potentials, Nature Phys 4, 213–217 (2008) DOI: 10.1038/nphys890

[13] Guang Bian et al, Experimental observation of two massless Dirac-fermion gases in graphene-topological insulator heterostructure, 2D Mater. 3, 021009 (2016) DOI: 10.1088/2053-1583/3/2/021009

[14] M. Park, Y. Kim and H. Lee, Design of 2D massless Dirac fermion systems and quantum spin Hall insulators based on sp-sp2 carbon sheets, Comp. Mater. 4, 54 (2018) DOI: 10.1038/s41524-018-0113-8

[15] C. Kamal, Massless Dirac-Fermions in Stable Two-Dimensional Carbon-Arsenic Monolayer, Phys. Rev. B 100, 205404 (2019) DOI: 10.1103/PhysRevB.100.205404

[16] N. P. Armitage, E. J. Mele and A. Vishwanath, Weyl and Dirac Semimetals in Three Dimensional Solids, Rev. Mod. Phys. 90, 015001 (2018) DOI: 10.1103/RevModPhys.90.015001

[17] D. Ciudad, Weyl fermions: Massless yet real, Nat. Mater. 14, 863 (2015) DOI: 10.1038/nmat4411

[18] Su-Yang Xu et al., Discovery of a Weyl fermion semimetal and topological Fermi arcs, arXiv:1502.03807 [cond-mat.mes-hall], Science 349, 613-617 (2015) DOI:10.1126/science.aaa9297

[19] H. -H. Lai, S. E. Grefe, S. Paschen and Q. Si, Weyl-Kondo semimetal in heavy-fermion systems, P. Natl. Acad. Sci. USA 115, 93-97 (2018) DOI: 10.1073/pnas.1715851115

[20] W. Yánez-Parreño, Y. S. Huang, S. Ghosh, et al., Thin film growth of the Weyl semimetal NbAs, Phys. Rev. Materials 8, 034204 (2024) DOI: 10.1103/PhysRevMaterials.8.034204

[21] Q. Lu, P. V. S Reddy, H. Jeon et al., Realization of a two-dimensional Weyl semimetal and topological Fermi strings, Nat Commun 15, 6001 (2024) DOI: 10.1038/s41467-024-50329-6

[22] J. Ikeda, K. Fujiwara, J. Shiogai et al., Critical thickness for the emergence of Weyl features in $Co_3Sn_2S_2$ thin films, Commun. Mater. 2, 18 (2021) DOI: 10.1038/s43246-021-00122-5

[23] X. Liu, S. Fang, Y. Fu, et al., Magnetic Weyl Semimetallic Phase in Thin Films of $Eu_2Ir_2O_7$, Phys. Rev. Lett. 127, 277204 (2021) DOI: 10.1103/PhysRevLett.127.277204

[24] T. Matsuda, N. Kanda, T. Higo, et al., Room-temperature terahertz anomalous Hall effect in Weyl antiferromagnet $Mn_3Sn$ thin films, Nat. Commun. 11, 909 (2020) DOI: 10.1038/s41467-020-14690-6

[25] A. Bedoya-Pinto, A. K. Pandeya, D. Liu, et al., Realization of Epitaxial NbP and TaP Weyl Semimetal Thin Films, ACS Nano 14, 4405-4413 (2020) DOI: 10.1021/acsnano.9b09997

[26] N. Yadav and N. Deo, Thin film Weyl semimetals with turning number of Fermi surface greater than unity, Physica E 158, 115901 (2024) DOI: 10.1016/j.physe.2024.115901

[27] C. W. Jang, Y. A. Salawu, J. H. Kim, et al., 2D Weyl-Semimetal States Achieved by a Thickness-Dependent Crossover and Topological Phase Transition in $Bi_{0.96}Sb_{0.04}$ Thin Films, Adv. Funct. Mater. 33, 2305179 (2023) DOI: 10.1002/adfm.202305179

[28] S. Jia, S. Y. Xu and M. Hasan, Weyl semimetals, Fermi arcs and chiral anomalies, Nature Mater 15, 1140–1144 (2016) DOI: 10.1038/nmat4787

[29] C. Guo, V. S. Asadchy, B. Zhao et al., Light control with Weyl semimetals. eLight 3, 2 (2023) DOI: 10.1186/s43593-022-00036-w

[30] Z. Jalali-Mola and S. A. Jafari, Electrodynamics of tilted Dirac and Weyl materials: A unique platform for unusual surface plasmon polaritons, Phys. Rev. B 100, 205413 (2019) DOI: 10.1103/PhysRevB.100.205413

[31] M. Fruchart, S. Jeon, K. Hur, V. Cheianov, U. Wiesner and V. Vitelli, Soft self-assembly of Weyl materials for light and sound, Proc. Natl. Acad. Sci. U.S.A. 115 (16) E3655-E3664 (2018) DOI: 10.1073/pnas.1720828115

[32] N. Nagaosa, T. Morimoto and Y. Tokura, Transport, magnetic and optical properties of Weyl materials, Nat Rev Mater 5, 621–636 (2020) DOI: 10.1038/s41578-020-0208-y

[33] K. Nakazawa, Y. Kato and Y. Motome, Magnetic, transport and topological properties of Co-based shandite thin films, *Commun Phys* 7, 48 (2024) DOI: 10.1038/s42005-024-01534-8

[34] J. P. Santos Pires, S. M. João, Aires Ferreira, B. Amorim and J. M. Viana Parente Lopes, Anomalous Transport Signatures in Weyl Semimetals with Point Defects, Phys. Rev. Lett. 129, 196601 (2022) DOI: 10.1103/PhysRevLett.129.196601

[35] Z. Long, Y. Wang, M. Erukhimova, M. Tokman and A. Belyanin, Magnetopolaritons in Weyl Semimetals in a Strong Magnetic Field, Phys. Rev. Lett. 120, 037403 (2018) DOI: 10.1103/PhysRevLett.120.037403

[36] P. Wang, Relativistic model of spontaneous wave-function localization induced by non-Hermitian colored noise, arXiv:2501.07050 [quant-ph] DOI: 10.48550/arXiv.2501.07050

[37] H. S. Choi, Fractionally charged Weyl spinors as the bases for elementary particles, arXiv:2503.15740 [physics.gen-ph] DOI: 10.48550/arXiv.2503.15740

[38] M. Thomson, Modern Particle Physics, Cambridge University Press, Cambridge (2013) ISBN: 9781107034266

[39] P. B. Pal, Dirac, Majorana, and Weyl fermions, Am. J. Phys. 79, 485–498 (2011) DOI: 10.1119/1.3549729

[40] D. H. Kobe, Gauge Invariance and the Dirac Equation, Int. J. Theor. Phys. 21, 685-702 (1982)

[41] D. J. Griffiths, Introduction to Electrodynamics (4th ed.), Cambridge University Press, Cambridge (2017) ISBN: 9781108420419

[42] M. Wakano, Intensely Localized Solutions of the Classical Dirac-Maxwell Field Equations, Progress of Theoretical Physics 35, 1117 – 1141 (1966) DOI: 10.1143/PTP.35.1117

[43] J Stubbe, Existence of localised solutions of (1+1)-dimensional nonlinear Dirac equations with scalar self-interaction, J. Phys. A: Math. Gen. 19, 3223 (1986), DOI: 10.1088/0305-4470/19/16/020

[44] C. J. Radford, Localized solutions of the Dirac–Maxwell equations, J. Math. Phys. 37, 4418–4433 (1996) DOI: 10.1063/1.531663

[45] H. Chaachoua Sameut, M. Asad-uz-zaman, U. Al Khawaja, et al., Families of Localized and Oscillatory Solutions to the Coupled-Nonlinear Two-Dimensional Dirac Equations. Phys. Wave Phen. 26, 306–311 (2018) DOI: 10.3103/S1541308X1804009X

[46] S. B. Faruque, S.D. Shuvo and P.K. Das, Localized states of Dirac equation, arXiv:1902.06232 [quant-ph] DOI: 10.48550/arXiv.1902.06232

[47] I. Bialynicki-Birula and Z. Bialynicka-Birula, Twisted localized solutions of the Dirac equation: Hopfionlike states of relativistic electrons, Phys. Rev. A 100, 012108 (2019) DOI: 10.1103/PhysRevA.100.012108

[48] G. N. Borzdov, Localized solutions of the Dirac equation in free space and electromagnetic space-time crystals, Phys. Rev. A 101, 012112 (2020) DOI: 10.1103/PhysRevA.101.012112